\documentclass[11pt,a4paper]{article}
\pdfoutput=1
\usepackage{jcappub}
\usepackage{graphicx, epsfig}
\usepackage{color}
\usepackage{mathrsfs}
\usepackage{bm}
\usepackage{amsmath,amssymb,yhmath}
\usepackage[caption=false]{subfig}
\usepackage{hyperref}
\usepackage[normalem]{ulem}
\usepackage{mathtools}

\newcommand{\be}{\begin{equation}}
\newcommand{\ee}{\end{equation}}
\newcommand{\ba}{\begin{eqnarray}}
\newcommand{\ea}{\end{eqnarray}}

\newcommand{\bfp}{{\bf p}}
\newcommand{\bfx}{{\bf x}}

\newcommand{\bfa}{{\bf a}}

\newcommand{\rmtr}{{\rm Tr}}
\newcommand{\calC}{{\cal C}}

\begin{document}

\title{Classical-Quantum Correspondence for Fields}

\author[a,b]{Tanmay Vachaspati}
\author[a,b]{George Zahariade}

\affiliation[a]{Physics Department, Arizona State University, Tempe, AZ 85287, USA}
\affiliation[b]{Beyond Center for Fundamental Concepts in Science, Arizona State University, Tempe, AZ 85287, USA}

\emailAdd{tvachasp@asu.edu}
\emailAdd{zahariad@asu.edu}

\abstract{We map the quantum problem of a free bosonic field in a space-time dependent background into a classical 
problem. $N$ degrees of freedom of a real field in the quantum theory are mapped into $2N^2$
classical simple harmonic oscillators with specific initial conditions. We discuss how this classical-quantum 
correspondence (CQC) may be used to evaluate quantum radiation and 
fully treat
the backreaction 
of quantum fields on classical backgrounds. 
The technique has widespread application, including to the quantum evaporation of classical breathers (``oscillons'').
}


\maketitle

\section{Introduction}

It has been known for some time that a quantum simple harmonic oscillator in one dimension
can be solved in terms of a classical simple harmonic oscillator in two dimensions
\cite{Lewis:1968yx, Lewis:1968tm, doi:10.1119/1.1986050,1969JETP...30..910P}. 
This mapping holds 
even if the parameters of the simple harmonic oscillator are time-dependent and provides a simple
method to calculate the quantum excitations of the oscillator due to a time varying frequency.
In \cite{Vachaspati:2018pps,Vachaspati:2018llo} we have developed this classical-quantum
correspondence (CQC) further and used it as an instrument to obtain the backreaction of the 
quantum excitations on the classical background. Comparison of the backreaction calculated
using the CQC to the backreaction calculated in a full quantum analysis for a simple
system -- a particle acted on by a constant force -- shows excellent agreement. Indeed, the
dynamics found using the CQC becomes more accurate as the background in the full quantum 
analysis becomes more classical.

Similar techniques have extensively been used to compute quantum particle production rates in time-dependent classical backgrounds~\cite{Zeldovich:1971mw, PhysRevD.9.3263, PhysRevD.12.368}. However the backreaction of this effect on the background has not been fully taken into account. In gravitational contexts, where renormalization of the energy-momentum tensor is required~\cite{HU1977217, Hu:1978zd}, it has only been computed to first iterative order in the semi-classical approximation. There, the particle production is calculated on a fixed background and its quantum average value is used as a source for the perturbations of said background. Technically this procedure should be reiterated {\it ad infinitum} but for computational complexity reasons only the first iteration is used. In the context of Schwinger pair production~\cite{PhysRevD.40.456, PhysRevLett.67.2427} the backreaction has been more fully addressed but mostly for homogeneous backgrounds. A notable exception is the work of Aarts and Smit~\cite{Aarts:1999zn} which uses the so-called ``mode function" method to discuss backreaction in full generality (see also~\cite{Borsanyi:2007wm,Saffin:2014yka}). 
Our aim is to show how the CQC can yield a technique, ultimately equivalent to the mode function method, allowing for the direct study of backreaction of quantum radiation on generic (homogeneous or inhomogeneous) classical backgrounds. In Appendix~\ref{proof} we have included a comparison between the CQC and standard semiclassical methods. 

The main focus of this paper is to extend the CQC to fields. We have in mind a system with a free quantum
field, $\phi$, that propagates in the background of a second classical field, $\Phi (x)$, or in a 
spacetime metric, $g_{\mu\nu}(x)$. In the first case an example of the action for $\phi$ is,
\be
S = \int d^4 x \left [ \frac{1}{2}  (\partial_\mu\phi)^2 - \frac{1}{2} \Phi^2 \phi^2 \right ]
\label{qS1}
\ee
while in the second case we may write,
\be
S = \int d^4 x \sqrt{-g} \frac{1}{2} g^{\mu\nu} \partial_\mu\phi \partial_\nu \phi .
\label{qS2}
\ee
The first example is relevant to cosmological inflation and phase transitions, while the second 
example is relevant to quantum radiation during gravitational collapse and cosmology. In this paper we will focus on the non-gravitational case given by the first example. This is particularly relevant to the study of the evaporation of soliton-like objects such as sine-Gordon breathers under the effect of quantum radiation. A numerical analysis of this phenomenon with the methods introduced in the current paper will be presented in a companion article~\cite{Olle:2019skb}.

We can convert the field theory problem to a quantum mechanics problem by discretizing the
action. One way is to define all the fields on a spatial lattice. In that case, the variables are
$\phi_{ijk}(t)$ where $ijk$ refer to a particular lattice site. Another way to discretize the action
is to expand all the fields in a complete basis of functions. In either case, the discrete action
is quadratic in the discrete variables and can be written generally as
\be
S_{\rm discrete} = \sum_{K,L} \left [ \frac{1}{2} {\dot \phi}_{K} {\cal M}_{KL} {\dot \phi}_{L} -
\frac{1}{2} \phi_{K} {\cal N}_{KL} \phi_{L} \right ]
\ee
where ${\cal M}$ and ${\cal N}$ denote time-dependent symmetric matrices and
subscripts $K$, $L$ denote generalized indices. Note that $\phi_K$ is only a function
of time.
Thus our problem reduces to an infinite set of quantum simple harmonic oscillators with
general time-dependent mass and spring-constant matrices. In practice we
will need to truncate the number of modes or consider a finite lattice and so we are
left with some large but finite number $N$ of quantum variables.

Having mapped the field theory to quantum mechanics we will now focus on the
solution of the quantum mechanics problem. In Sec.~\ref{quantumsystem} we set up
the quantum problem of $N$ simple harmonic oscillators with time dependent mass
matrices and frequencies. We solve the Heisenberg equations for the ladder
operators and provide a classical-quantum mapping  in Sec.~\ref{heisenberg}.
We discuss constraints and count the independent degrees of freedom in 
Sec.~\ref{constraints}. In Sec.~\ref{energy} we show the key result that the 
expectation of the quantum Hamiltonian equals the energy of the classical oscillators. 
In Sec.~\ref{clFT} we consider if the classical system can be written as a 
classical field theory and we conclude in Sec.~\ref{conclusions}. In Appendix~\ref{proof}, we further provide a proof for why the CQC gives the correct backreaction. We also compare it to standard semiclassical methods.

\section{The quantum system}
\label{quantumsystem}

We consider $N$ coupled simple harmonic oscillators whose quantum dynamics are fully described by the Hamiltonian
\be
H = \frac{1}{2} \bfp^T \mu^{-2} \bfp + \frac{1}{2} \bfx^T \mu \Omega^2 \mu \bfx\ ,
\label{ham}
\ee
where $\bfp = (p_1,\ldots,p_N)^T$ are the momentum operators corresponding to the position 
operators $\bfx=(x_1,\ldots,x_N)^T$, and ${}^T$ denotes matrix transposition. The matrices 
$\mu=[\mu_{ij}]_{1\leq i,j\leq N}$ and $\Omega=[\Omega_{ij}]_{1\leq i,j\leq N}$ are assumed 
to be real and symmetric positive definite and can depend on time. Note that here and henceforth 
we employ the matrix notation but, since matrix elements need not commute with each other, 
expressions must be handled with care. In particular usual matrix identities such as $(AB)^T= B^TA^T$ 
do not necessarily hold if $A$,$B$ are operator valued matrices.

We can define ladder operators $\bfa^\dag=(a_1^\dag,\ldots,a_N^\dag)$ and $\bfa =(a_1,\ldots,a_N)^T$ 
via the usual procedure
\ba
\label{ladder}
\bfa &=& \frac{1}{\sqrt{2}}\left(\sqrt{\Omega}^{-1}\mu^{-1} \bfp  - i  \sqrt{\Omega} \mu \bfx\right)\ ,\\
\bfa^\dag &=& \frac{1}{\sqrt{2}}\left(\bfp^T\mu^{-1}\sqrt{\Omega}^{-1} + i \bfx^T \mu \sqrt{\Omega}\right)\ . 
\ea
Some care is required for the proper understanding of the generalized adjoint operator ${}^\dag$ . 
Indeed, taking the adjoint of a matrix means first transposing it and then taking the Hermitian conjugate 
of its elements, the latter operation reducing to a mere complex conjugation when the entries are $c$-numbers. 
Note also that $\sqrt{\Omega}$ is defined in the usual way by first diagonalizing $\Omega$ by
a similarity transformation, then taking the positive square root of the resulting diagonal matrix, and finally 
performing the inverse similarity transformation.
  
The Hamiltonian can then be rewritten in terms of ladder operators as
\be
H = \bfa^\dag\Omega \bfa + \frac{1}{2} {\rm Tr}(\Omega)\ .
\label{hamaa}
\ee
Indeed we can check this by straightforward multiplication since
\be
\label{hamaadag}
\bfa^\dag\Omega \bfa = H
+ \frac{i}{2}
\left [ 
\bfx^T \mu \Omega \mu^{-1} \bfp - \bfp^T\mu^{-1} \Omega \mu \bfx
\right ]\ ,
\ee
and the second term is evaluated to be $i {\rm Tr}(\Omega)$ by using
the symmetry of $\mu$ and $\Omega$ as well as the commutation
relation $[x_i,p_l]=i\delta_{il}$.

To find the quantum dynamics of this system, we work in the Heisenberg picture from now on. Since the ladder operators verify
\be
[a_i, a^\dag_j] = \delta_{ij}, \ \ [a_i, a_j] = 0 = [a^\dag_i, a^\dag_j]\ ,
\label{comm}
\ee
and
\be
[\bfa, H] = \Omega \bfa , \ \ [\bfa^\dag,H] = -\bfa^\dag \Omega\ ,
\ee
the Heisenberg equations are
\ba
\frac{d\bfa}{dt} &=& -i \Omega \bfa + \frac{\partial\bfa}{\partial t}\ , \\
\frac{d\bfa^\dag}{dt} &=& + i \bfa^\dag \Omega + \frac{\partial\bfa^\dag}{\partial t}\ ,
\ea
where the partial time derivatives should be understood as acting only on the explicitly time dependent part of the operators, 
{\it i.e.}
\ba
\frac{\partial\bfa}{\partial t}&=&\frac{1}{2}\bigg[\frac{d}{dt}\left(\sqrt{\Omega}^{-1}\mu^{-1}\right)\mu\sqrt{\Omega}\left(\bfa+\bfa^\dag{}^T\right)
-\frac{d}{dt}\left(\sqrt{\Omega}\mu\right)\mu^{-1}\sqrt{\Omega}^{-1}\left(\bfa^\dag{}^T-\bfa\right)\bigg]\ ,
\label{derladder}
\\
\frac{\partial\bfa^\dag}{\partial t}&=&\frac{1}{2}\bigg[\left(\bfa^T+
\bfa^\dag\right)\sqrt{\Omega}\mu\frac{d}{dt}\left(\mu^{-1}\sqrt{\Omega}^{-1}\right) -\left(\bfa^T-\bfa^\dag\right)\sqrt{\Omega}^{-1}\mu^{-1}\frac{d}{dt}\left(\mu\sqrt{\Omega}\right)\bigg]\ .
\ea

\section{Map to the classical system}
\label{heisenberg}

To solve these equations we follow the same procedure as in Ref.~\cite{Vachaspati:2018llo} 
and introduce the Bogoliubov coefficient matrices $\alpha=\left[\alpha_{ij}(t)\right]_{1\leq i,j\leq N}$ 
and $\beta=\left[\beta_{ij}(t)\right]_{1\leq i,j\leq N}$ defined by
\ba
\bfa &=& \alpha\ \bfa_{0} + \beta\ \bfa_0^\dag{}^T, \\ 
\bfa^\dag &=& \bfa_0^\dag\ \alpha^\dag+\bfa_0^T\ \beta^\dag,
\label{AAdag}
\ea
where the $0$ subscript refers to the operators at the initial time. The transposition operation in the last terms of these equations is a necessity given our initial definitions for $\bfa$ and $\bfa^\dag$ as column and row vectors respectively. The commutation relations \eqref{comm} imply the existence of the constraint equations  
\ba
&&
\alpha\beta^T-\beta\alpha^T=0,
\label{norm1} \\ &&
\alpha\alpha^\dag-\beta\beta^\dag=1.
\label{norm2}
\ea

The Heisenberg equations then imply that $\alpha$ and $\beta$ satisfy
\ba
\frac{d\alpha}{dt} &=& - i \Omega {\alpha} +\frac{1}{2}\bigg[\frac{d}{dt}\left(\sqrt{\Omega}^{-1}\mu^{-1}\right)\mu\sqrt{\Omega}\left(\alpha+\beta^{*}\right) 
-\frac{d}{dt}\left(\sqrt{\Omega}\mu\right)\mu^{-1}\sqrt{\Omega}^{-1}\left(\beta^{*}-\alpha\right)\bigg],\label{alphaeom} \\
\frac{d\beta}{dt} &=& - i \Omega {\beta} +\frac{1}{2}\bigg[\frac{d}{dt}\left(\sqrt{\Omega}^{-1}\mu^{-1}\right)\mu\sqrt{\Omega}\left(\alpha^{*}+\beta\right)
-\frac{d}{dt}\left(\sqrt{\Omega}\mu\right)\mu^{-1}\sqrt{\Omega}^{-1}\left(\alpha^{*}-\beta\right)\bigg],\label{betaeom} 
\ea
with initial conditions $\alpha=1$ and $\beta=0$. The particular form of these equations suggests the following change of variables
\ba
P&=&\frac{1}{\sqrt{2}}\mu\sqrt{\Omega}(\alpha^{*}+\beta)\ ,\\
iZ&=&\frac{1}{\sqrt{2}}\mu^{-1}\sqrt{\Omega}^{-1}(\alpha^{*}-\beta)\ ,
\ea
or
\ba
\alpha&=& \frac{1}{\sqrt{2}}\left(\sqrt{\Omega}^{-1}\mu^{-1} P^*  - i  \sqrt{\Omega} \mu Z^*\right)\ ,
\label{alphaPZ} \\
\beta&=&\frac{1}{\sqrt{2}}\left(\sqrt{\Omega}^{-1}\mu^{-1} P  - i  \sqrt{\Omega} \mu Z\right)\ .
\label{betaPZ}
\ea
Indeed in these variables the equations simplify significantly, reducing to
\be
\dot{P}=-\mu\Omega^2\mu Z\quad\text{and}\quad\dot{Z}=\mu^{-2}P\ ,
\label{PZeom}
\ee
while the initial conditions become 
\be
P_0=\frac{1}{\sqrt{2}}\mu_0\sqrt{\Omega_0}\quad\text{and}\quad Z_0=-\frac{i}{\sqrt{2}}\mu_0^{-1}\sqrt{\Omega_0}^{-1}\ .
\label{PZic}
\ee
Here and henceforth we use the usual dot notation to represent time derivatives since there is no ambiguity left 
between partial and total derivatives.
The equations of motion for $Z$, $P$ can be derived from the classical Hamiltonian,
\be
H_c =  \frac{1}{2} \rmtr \left [ P^\dag \mu^{-2} P + Z^\dag \mu \Omega^2 \mu Z \right ],
\label{Hc}
\ee
which is simply a rewrite of the original Hamiltonian for $\bfx$, $\bfp$ in \eqref{ham} in terms of the new
variables $Z$, $P$.

\section{Constraints and conserved quantities}
\label{constraints}

We can check that the constraints in \eqref{norm1} and \eqref{norm2} are consistent with the
evolution equations \eqref{alphaeom} and \eqref{betaeom}. Thus if the constraints are satisfied
at the initial time then they continue to hold as the system evolves and no secondary constraints arise.
The constraints can also be rewritten in terms of $P$ and $Z$ using \eqref{alphaPZ} and \eqref{betaPZ}
as,
\ba
{\cal C}_1 &\equiv& P^*P^T-PP^\dag = 0, \label{calC1}\\ 
{\cal C}_2 &\equiv& Z^*Z^T-ZZ^\dag = 0, \label{calC2}\\ 
{\cal C}_3 &\equiv& \ \ i(ZP^\dag-Z^*P^T) = 1.\label{calC3}
\label{PZconstraints}
\ea
Since $P$ and $Z$ are simply another way of writing $\alpha$ and $\beta$, the constraints in terms
of $P$ and $Z$ are also consistent with the evolution equations. Note that the constraints written in 
this form are either purely imaginary or real, as opposed to the original ones written in terms of $\alpha$ 
and $\beta$ which were complex. This, along with their symmetry properties, explains why we now need 
three matrix equations instead of two to write down the same number of constraints.

There are $2N^2$ real components of $Z$ and also of $P$. This suggests that there are a total
of $4N^2$ real degrees of freedom. However, this is not correct because the constraints relate 
different components of $Z$ and $P$, although in a complicated way.

Consider the matrix $\alpha\beta^T-\beta\alpha^T$. This is antisymmetric as can be checked by
taking the transpose. Therefore it only has $N(N-1)/2$ independent complex entries, or 
$N(N-1)$ real entries. So \eqref{norm1} 
provides $N(N-1)$ constraints on the $4N^2$ total number of real numbers in $\alpha$ and $\beta$.
Next we consider the matrix $\alpha\alpha^\dag-\beta\beta^\dag$. Since this matrix is Hermitian
it has $N^2$ independent real components and \eqref{norm2} provides
$N^2$ constraints. Hence the independent (real) degrees of freedom of the matrices 
$\alpha$ and $\beta$ are given by
\be
4N^2 - N(N-1) - N^2 = 2N^2 + N.
\label{constraineddof}
\ee
Since $\alpha$, $\beta$ and $Z$, $P$ are related by a linear
transformation, the number of independent real degrees of freedom in $Z$, $P$
are in general also $2N^2+N$. A full analysis of the constraint structure ``\`a la Dirac'' 
indeed shows that none of the above constraints are first-class and thus the number of 
degrees of freedom cannot be reduced further.

In addition to the constraints, the evolution equations also have some conserved
quantities. The difference between constraints and conserved quantities is that
the constraints are satisfied during evolution only if they are satisfied initially, while
the conservation of quantities holds irrespective of the initial conditions. This
may be illustrated for say $Z^*Z^T-ZZ^\dag=0$ constraint in \eqref{calC2}.
\be
{\dot {\cal C}_2}= (Z^*P^T-ZP^\dag)\mu^{-2}-\mu^{-2}(Z^*P^T-ZP^\dag)^T.
\nonumber
\ee
If we now use the value of $Z^*P^T-ZP^\dag$ from \eqref{PZconstraints}, we see
that the right-hand side vanishes and the $Z^*Z^T-ZZ^\dag=0$ constraint continues 
to hold with time. (This is in fact a manifestation of the fact that there are no secondary 
constraints.) On the other hand, the system has two conserved quantities
\ba
J &=& i (P^\dag Z- Z^\dag P), 
\label{J} \\ 
{\bar J}&=&i(P^\dag Z^* - Z^\dag P^*).
\label{barJ}
\ea
The conservation of $J$ and ${\bar J}$ holds independently of their initial values
as is seen by checking ${\dot J}=0={\dot {\bar J}}$. 
However, the conservation of $J$ and ${\bar J}$ cannot be used to further
limit the number of degrees of freedom because of the relations
\ba
Z J + Z^* {\bar J}^*  &=& +i {\cal C}_2 P + {\cal C}_3 Z, \\
P J + P^* {\bar J}^* &=& -i{\cal C}_1 Z + {\cal C}_3^\dag P .
\ea
Since $Z$, $P$ satisfy the constraints, we have $\calC_1=0=\calC_2$
and $\calC_3=1$. This leads to $J=1$ and ${\bar J}=0$ which is
consistent with the initial conditions in \eqref{PZic}. Hence we are
still left with the $2N^2+N$ degrees of freedom.

This degree of freedom counting actually hides a symplectic structure since the quantum evolution 
of the ladder operators is given by the action of the symplectic group ${\rm Sp}(2N,\mathbb{R})$ 
whose dimension is $2N^2+N$. Indeed the matrices
\be
\left(\begin{array}{cc}\alpha & \beta \\ \beta^* & \alpha^*\end{array}\right)\quad{\rm and}\quad
\left(\begin{array}{cc}P^* & P \\ iZ^* & iZ\end{array}\right)
\ee
as well as their transpose can be shown to belong to a subgroup of  ${\rm Sp}(2N,\mathbb{C})$ 
isomorphic to ${\rm Sp}(2N,\mathbb{R})$ as a consequence of the previously discussed constraints.

In certain physical settings the problem can reduce further. For example, if
$\Omega$ is diagonal, we can check that $Z$ and $P$ are also diagonal. 
In this case, \eqref{norm2} provides $N$ constraints on the $2N+2N$ real components
of $Z$, $P$ for a total of $4N-N=3N$ real degrees of freedom. 

In practice, for example in a numerical implementation, it seems simpler to
solve the $4N^2$ equations for $Z$, $P$ instead of first reducing the system
to $2N^2+N$ degrees of freedom. 
The straight-forward
solution of the $Z$, $P$ equations is further simplified because the equations
of motion do not mix different columns of $Z$, $P$. Thus one could solve the system
column by column, say one per processor, each column having different initial conditions 
but identical differential equations. To make this more explicit, we can re-write 
\eqref{PZeom} as
\be
\dot{P}_i^{(j)}=-(\mu\Omega^2\mu)_{ik} Z_k^{(j)}
\quad\text{and}\quad\dot{Z}_i^{(j)}=\mu_{ik}^{-2}P_k^{(j)}\ ,
\label{PZeomcolumn}
\ee
where the superscript refers to the column. Thus the equations are independent of
$j$, though the initial conditions do depend on the column. The energy \eqref{Hc} too 
becomes a sum over the columns that we can write explicitly,
\be
H_c =  \sum_{j=1}^N \frac{1}{2} \left [ P_{i}^{(j)*} (\mu^{-2})_{ik} P_{k}^{(j)} 
+ Z_{i}^{(j)*} (\mu \Omega^2 \mu)_{ik} Z_{k}^{(j)} \right ].
\ee

\section{Energy}
\label{energy}

As is standard in the Bogoliubov approach, we
take the expectation value of \eqref{hamaa} in the (initial) vacuum state to find the
quantum energy of the simple harmonic oscillators
\be
E_q \equiv \langle 0 | H | 0 \rangle = \rmtr (\beta^\dag \Omega \beta) + \frac{1}{2}\rmtr (\Omega ).
\label{Eq}
\ee
Next we use \eqref{betaPZ} to obtain
\ba
\rmtr (\beta^\dag \Omega \beta) = 
E_c +\frac{i}{2} \rmtr \{ Z^\dag \mu \Omega \mu^{-1} P - P^\dag \mu^{-1} \Omega \mu Z \}\,,
\label{betabeta}
\ea
where $E_c$ is the energy in $Z$, $P$ as given by the Hamiltonian in \eqref{Hc}.
The second term in the above equation can be recast as
\be
\frac{i}{2}\rmtr \{ Z^\dag \mu \Omega \mu^{-1} P - P^\dag \mu^{-1} \Omega \mu Z \} = -\frac{1}{2}\rmtr \{\calC_3\mu^{-1}\Omega\mu\}\,,
\label{cross}
\ee
which by virtue of \eqref{calC3} is simply $-\rmtr (\Omega)/2$.
Inserting \eqref{cross} into \eqref{betabeta} and then combining with \eqref{Eq} leads to the 
key result
\be
E_q = E_c.
\ee
Therefore the quantum energy can be found directly as the classical energy in $Z$ and $P$. 
Notice that the associated classical Hamiltonian $H_c$ can be derived from the Lagrangian
\be
L_c=\frac{1}{2} \rmtr \left [ \dot{Z}^\dag \mu^{2} \dot{Z} - Z^\dag \mu \Omega^2 \mu Z \right ]\ ,
\label{classlag}
\ee
which is invariant under the transformation $Z\rightarrow ZU$ where $U$ is a constant $N\times N$ unitary matrix.
The model has a global $U(N)$ symmetry.

This completes our re-writing of the quantum dynamics of $N$ simple harmonic oscillators in terms 
of the solution for $2N^2$ classical simple harmonic oscillators with the specific initial conditions
given in \eqref{PZic}.

\section{From a quantum to a classical field theory?}
\label{clFT}

The question is if we can write the $Z$, $P$ system as a classical field theory. If so,
we would have mapped the original quantum field theory to a classical field theory.
This is simple to do if $\Omega$ is diagonal for then $Z$ and $P$ are also diagonal. Then
the diagonal elements of $Z$ can be thought of as the mode coefficients of a complex 
scalar field and $P$
their canonical momenta. In this case, the quantum real scalar field is mapped
to a classical complex scalar field and the initial conditions are such that the modes
carry a certain amount of energy and global charge as noted in \cite{Vachaspati:2018pps}.
Can a similar mapping be made for general $\Omega$?

For concreteness, let us discuss the example of a free massless quantum scalar field $\phi$ in a classical background 
$\Phi$ alluded to in Eq. \eqref{qS1}. For added simplicity we restrict ourselves to the $1+1$ dimensional case. The 
relevant action will be
\be
S_\phi=\frac{1}{2}\int dt dx\left[\dot{\phi}^2-\phi'^2-\lambda\Phi ^2\phi^2\right]\,, 
\label{Sphi}
\ee
where an overdot and a prime denote partial differentiation with respect to $t$ and $x$ respectively,
and $\lambda$ is a coupling constant with dimensions of inverse length squared. Notice that 
in 1+1 dimensions the scalar fields have mass dimension equal to zero.

We first discretize \eqref{Sphi} 
by putting it on a spatial lattice with $N$ sites spaced by a distance 
$a$. For any integer $i$ running from $1$ to $N$ we define
\ba
\Phi(t,ia)&=&\Phi_i (t)\,,\\
\phi(t,ia)&=&\phi_i(t)\,,\\
\phi''(t,ia)&=&\frac{1}{a^2}(\phi_{i+1}(t)-2\phi_i(t)+\phi_{i-1}(t))\label{secderdisc}\,.
\ea 
We will impose Dirichlet boundary conditions $\phi_0=\phi_{N+1}=0$ at ``spatial infinity." 
In order to be able to use the results of the previous sections, we further define $\mathbf{x}(t)$ to be 
the column vector $(a\phi_1,\dots,a\phi_{N})^T$.
With these conventions the discretized action will read 
\be
S_\phi\approx\int dt\ \frac{1}{a}\left[\frac{1}{2}\dot{\mathbf{x}}^T\dot{\mathbf{x}}
-\frac{1}{2}\mathbf{x}^T\Omega^2\mathbf{x}\right]\,,
\ee
where
\be
\Omega^2_{ij}=
\begin{cases}
2/a^2+\lambda\Phi_i^2\,,\ \text{if}\ i=j\\
-1/a^2\,,\ \text{if}\ i=j\pm 1 \,.
\end{cases}
\label{Omega2}
\ee
This action obviously yields a Hamiltonian of the form \eqref{ham} (with the matrix $\mu$ 
being $1/\sqrt{a}$ times the identity). Therefore, according to Eq. \eqref{classlag}, the dynamics of the quantum 
degrees of freedom in $\mathbf{x}$ will be described by the classical action 
\be
S_c=\int dt\frac{1}{2a}\rm{Tr}\left[\dot{Z}^\dag \dot{Z}-Z^\dag\Omega^2Z\right]
\label{Sc}
\ee 
where $Z$ is a complex matrix obeying the initial conditions
\be
Z_0=-i\sqrt{\frac{a}{2}}\sqrt{\Omega_0}^{-1}\quad\text{and}\quad\dot{Z}_0=\sqrt{\frac{a}{2}}\sqrt{\Omega_0}\,.
\label{eqinitial}
\ee
It should be noted that the choice $\Phi(t=0)$, where $t=0$ is the initial time,
is crucial and defines the vacuum for the quantum field $\phi$.
We can also start the evolution at different snapshots (initial times) of some chosen background
and that will lead to different evolutions. 
This corresponds to the vacuum ambiguity, for example of de Sitter space~\cite{Danielsson:2002kx}.
Given the particular form of the matrix $\Omega^2$ in \eqref{Omega2} and provided that 
\be
a^2\ll\frac{1}{\lambda\Phi_i^2}\label{approx0}\,,
\ee
it is straighforward to show that
$\Omega_0^2=ODO^{T}$,
where 
\be
D_{ij}=\frac{4}{a^2}\sin^2\left(\frac{\pi i}{2(N+1)}\right)\delta_{ij}\label{approx2}\,,
\ee
and $O$ is an orthogonal matrix with components,
\be
O_{ij}=\sqrt{\frac{2}{N+1}}\sin\left(\frac{\pi ij}{N+1}\right)\label{approx3}\,.
\ee

Note that the classical action in \eqref{Sc} is the discretization of
\be
S_\psi=\frac{1}{2}\int dt\,dx\,dy \Big[|\dot{\psi}(t,x,y)|^2-|\psi'(t,x,y)|^2 -\lambda\Phi(t,x)^2|\psi(t,x,y)|^2\Big],
\label{disc}
\ee
where the field $\psi$ is a complex scalar field defined over twice the number of spatial dimensions
and $Z_{ij}$ corresponds to $a^{3/2}\psi(t,ia,ja)$. 
The form of the classical equation of motion for $\psi$ will thus be identical to that of 
$\phi$ but in finding solutions, we have to keep in mind that the initial conditions
for $\psi (t,x,y)$ can depend non-trivially on $y$.

For practical purposes the discretized action will be most useful. But in order to have a 
fully consistent picture we need to understand the intricacies of the large $N$ and small $a$ 
limits, or how the program of renormalization, inherently present in any field theory, carries over 
to this discretized classical action.

In the following, we will consider the limit $N\rightarrow\infty$ and $a\rightarrow 0$ while the physical 
size of the lattice $L=a(N+1)$ is held fixed. This continuum limit is most relevant for numerical calculations 
related to particle physics where a hard UV cutoff is not necessarily physical. Notice that in the context of 
condensed matter theory, the so-called {\it thermodynamic} limit $N\rightarrow\infty$, $L\rightarrow\infty$ 
while the lattice spacing $a=L/(N+1)$ is held fixed, would be more relevant.

We will make use of Eqs. \eqref{approx2} and \eqref{approx3} whose validity is ensured since $a$ vanishes 
in the large $N$ limit (and \eqref{approx0} is satisfied), to estimate the asymptotic behavior of various physical 
quantities.

A first divergence arises when examining the energy of the system at time $t=0$. Indeed the quantity
\be
E_q=E_c=\frac{1}{2}\text{Tr}(\Omega_0)\sim \frac{2N^2}{\pi L}\sim\frac{2L}{\pi a^2}\,,
\ee
diverges in the continuum limit as is expected, since it is the zero-point value of the energy of a system 
with an infinite number of degrees of freedom. However since only relative energies are measurable in 
this setup, we can subtract off the zero-point energy. 
The renormalization of the stress tensor
for gravitational systems as described in Ref.~\cite{Birrell:1982ix} for example will also have a 
corresponding procedure in the CQC. In this paper we restrict ourselves to non-gravitational field theory.

Other divergences appear when the background $\Phi$ is given its own dynamics through the action
\be
S_{\Phi}=\int dt dx\left[\frac{1}{2}\dot{\Phi}^2-\frac{1}{2}\Phi'^2-V(\Phi)\right]\,,
\ee
where $V$ is a generic potential, and thus backreaction is taken into account. In this case the system of 
discretized equations reads,
\ba
&&
\hspace{-0.25in}
\ddot{Z}_{ij}+\Omega^2_{ik}Z_{kj}=0\label{eqclass}\,,\\
&&
\hspace{-0.25in}
\ddot{\Phi}_i-\frac{1}{a^2}\left(\Phi_{i+1}-2\Phi_i+\Phi_{i-1}\right)+V'(\Phi_i)
+\lambda \left(\frac{1}{a^2}\sum_{j=1}^NZ^*_{ij}Z_{ij}\right)\Phi_i=0\label{eqback}\,,
\ea
where we have used prescription \eqref{secderdisc} to discretize the spatial derivative of the background field. 
At time $t=0$, the term in brackets in Eq. \eqref{eqback} diverges as,
\be
\frac{1}{a^2}\sum_{j=1}^NZ^*_{ij}Z_{ij} \biggr |_{t=0} =
\frac{1}{2a}\Omega_0^{-1}{}_{ii}\sim \frac{1}{2aN} \text{Tr}(\Omega_0^{-1})\sim \frac{1}{2\pi}\ln N,
\label{estimate}
\ee
where the index $i$ is {\it not} summed over. 

Now if we were to solve the equations for another value of $N$, say
$N'=\zeta N$, \eqref{estimate} tells us that the factor in parenthesis in
\eqref{eqback} shifts by $\ln(\zeta)/(2\pi)$. This shift is completely equivalent
to a shift in the classical mass of $\Phi$. In other words, rescaling $N$ by
a factor $\zeta$ is the same as shifting the classical potential $V(\Phi)$
by $ \lambda \ln(\zeta) \Phi^2/(4\pi)$, which is equivalent to 
renormalizing the mass of $\Phi$. In general, we then need an experimental
input that tells us what the physical mass is at a given resolution or energy
scale. In our case, we simply need to specify the classical potential at some
large value of $N=N_*$ (or some small value of $a=a_*$) and then take
the renormalized classical potential for any value of $N$ to be
\be
V_R(\Phi) = V(\Phi) - \frac{\lambda}{4\pi} \ln \left (\frac{N}{N_*} \right ) \Phi^2
\label{renpot}
\ee
Then the $N$ dependence of the renormalized potential will cancel the
$N$ dependence of the term in parenthesis in \eqref{eqback}. The
resulting physical equation of motion will be independent of $N$, apart
from discretization errors in evaluating the Laplacian term.

%

In terms of the renormalized potential Eq.~\eqref{eqback} reads
\be
\ddot{\Phi}_i-\frac{1}{a^2}\left(\Phi_{i+1}-2\Phi_i+\Phi_{i-1}\right)+V_R'(\Phi_i)
+\frac{\lambda}{a^2}\sum_{j=1}^NZ^*_{ij}Z_{ij}\Phi_i=0\,,
\label{eqbackbis}
\ee
where $V_R$ is given in \eqref{renpot}.


%

To summarize, the relevant equations of motion
are Eqs.~\eqref{eqclass} and \eqref{eqbackbis}, with initial conditions in \eqref{eqinitial}.
We thus have a consistent prescription that allows us to deal with 
UV divergences peculiar to quantum field theory in the CQC.

\section{Conclusions}
\label{conclusions}

We can summarize the CQC for fields as follows.
We are interested in the evolution of a free bosonic quantum field in a classical background.
The quantum field problem can be mapped to a quantum system of $N$ simple harmonic 
oscillators with time-dependent frequencies and masses (see \eqref{ham}) that start off in 
their ground state.
The CQC stipulates that this quantum dynamics can be evaluated entirely using a classical system
of $2N^2+N$ real variables; or more straight-forwardly as $2N^2$ simple harmonic oscillators 
($4N^2$ phase space variables) with specific initial conditions and $2N^2-N$ conserved quantities. 
The Hamiltonian for the $2N^2$ simple harmonic oscillators is given by \eqref{Hc} and,
crucially, the initial conditions for the classical evolution are given by \eqref{PZic}.

Next suppose that we have a model for the agency that is responsible for the time
dependence of the masses and frequencies of the simple harmonic oscillators. 
As discussed in the introduction, this could be due to the dynamics of a background field 
or the spacetime metric. We wish to obtain the backreaction of the quantum excitations on 
the background, but the background is classical while the excitations are quantum. 
And this is where the CQC can help since it maps the quantum problem into
a classical problem. Then a classical Hamiltonian can be written for the entire
system,
\be
H = H_\Phi + H(Z,P;\Phi)
\ee
where $\Phi$ denotes the classical background field and the Hamiltonian for
$Z$ and $P$ depends on this background but is also classical. Hence we can
solve the classical problem for $\Phi$, $Z$ and $P$ and this will be the desired
solution that includes backreaction. Note however, that although the equations of motion for 
the matrices $Z$ and $P$ do not directly couple different columns to each other (as mentioned 
at the end of Section \ref{constraints}), because the dynamics of $\Phi$ are sourced collectively 
by all these columns, one cannot solve the problem column by column anymore in the backreacting case.

Before closing we would like to highlight a few salient points. 
The CQC is an exact mapping from the quantum problem to the classical problem. Given the 
classical solution, we can reconstruct the quantum evolution in its entirety.
The CQC holds for {\it any} time dependence of the masses and frequencies of the original 
quantum problem. Then, even with the backreaction included, the CQC is exact, since
the backreaction simply modifies the time dependence of the masses and frequencies.
Departures from the CQC only occur if the background itself is not completely classical.
In the example of a particle acted on by a constant force~\cite{Vachaspati:2018llo} we found 
that the CQC becomes more accurate as the quantum spreading of the particle's wavepacket
becomes slower than the speed of rolling. Since the rolling speed grows with
time, the CQC becomes more accurate at late times.

For clarity, we contrast the CQC with the Wigner representation of 
quantum mechanics~\cite{PhysRev.40.749}.
In the CQC we are interested in quantum particles or fields in
classical backgrounds and the backreaction of those quantum fields
on the background. To formulate the solution of this problem, we have
found a general solution for the quantum variables in terms of a
solution to a classical problem, which then enables us to find the 
quantum backreaction on the classical background.

More concretely, if our quantum variable is a single simple
harmonic oscillator whose position is $x$, then the corresponding
classical variable, denoted by the complex function $z(t)$, is a
simple harmonic oscillator variable in two dimensions. 
The solution for the quantum problem, namely the wavefunction in the
Schr\"odinger representation or operators in the Heisenberg
representation, can be written in terms of the solution
to the classical equations of motion for $z(t)$ with specific
initial conditions.
The quantum backreaction on the classical background is then found
by simultaneously solving classical equations of motion for $z(t)$
and the background, again with specific initial conditions. 

In contrast, the Wigner representation defines a quasi-probability
function on phase space, $W(x,p,t)$, that contains the same information
as the wavefunction. So it is a classical formulation
of quantum dynamics but this is where the similarity with the CQC
ends. In particular, to apply the CQC we need only find $z(t)$ --
the trajectory of a particle in two dimensions -- whereas in the
Wigner representation we would need to solve for a function
on phase space $W(x,p,t)$. Further, the question of interest to
us, namely the quantum backreaction on classical backgrounds, is
not, to our knowledge, one that is addressed using the Wigner
representation.

Our analysis in this paper extends the CQC to the realm of quantum field theory
and can potentially be useful in a vast number of applications. 
In a companion paper~\cite{Olle:2019skb} we illustrate the backreaction analysis for an oscillon 
undergoing quantum evaporation.

\acknowledgments
TV is grateful to Bei-Lok Hu for comments and the Institute 
for Advanced Study, Princeton for hospitality.
TV's work is supported by the U.S. Department of Energy,
Office of High Energy Physics, under Award No.~DE-SC0019470 at Arizona State
University and GZ is supported by John Templeton Foundation grant 60253.

\appendix

\section{CQC and semiclassical approximation}
\label{proof}

In this appendix we elucidate the relationship between the CQC and the semiclassical approximation in the context of the class of field theory models described in the paper. This provides a justification for why the CQC gives the correct backreaction and shows how it improves upon standard methods.
Consider first the classical equation of motion for the background which for
illustration purposes we will take to be given by a scalar field,
\be
\square \Phi + V(\Phi ) + \lambda \phi^2 \Phi = 0.
\ee
The field $\Phi$ is to be treated classically while $\phi$ is considered quantumly.
Hence, in the semiclassical approximation, $\phi^2$ in the equation of motion is
replaced by its vacuum expectation value
\be
\square \Phi + V(\Phi ) + \lambda\, {{}_H \langle} 0| \phi_H^2 |0\rangle_H \Phi = 0,
\ee
where the $H$ subscripts on the vacuum state and $\phi$ emphasize that 
we are working in the Heisenberg picture where quantum states do
not evolve but operators do evolve. The key point now is that we
know the evolution of the operator $\phi_H^2$, and hence of $\langle \phi^2\rangle$,
in terms of the c-number variables $Z$. In our lattice formulation, this is
\be
{{}_H\langle} 0| \phi_H^2 |0\rangle_H \bigr |_{x=i a} 
= \frac{1}{a^2} \sum_{j=1}^N Z_{ij}[\Phi]^* Z_{ij}[\Phi]
\ee
provided $Z$ satisfies its own equation of motion with suitable initial conditions
as described above. Hence the background satisfies the equation of motion in
\eqref{eqback}. Together with the equation of motion for $Z$ in \eqref{eqclass}
that is valid for arbitrary backgrounds we obtain the full set of CQC equations.

The semiclassical approximation has been a tool for many years
and it is worthwhile to clarify how the CQC approach is different from existing
analyses. The semiclassial approximation is usually considered iteratively. 
At first one has a zeroth order solution for the classical background, call
it $\Phi^{(0)}$. Expectation values of the quantum operators are calculated
in this background, which we can denote $\langle \phi^2 \rangle^{(0)}$.
These are then inserted in the background equations to get a quantum
corrected background, $\Phi^{(1)}$ on which the corrected expectation
value $\langle \phi^2 \rangle^{(1)}$ can be calculated, and so on. At the
$n^{th}$ step of this iteration, the equation for the background looks
like
\be
\square \Phi^{(n)} + V(\Phi^{(n)} ) 
+ \lambda \langle 0| \phi^2 |0\rangle^{(n-1)} \Phi^{(n)} = 0,
\ee
and the Heisenberg equations for the quantum operator $\phi^{(n)}$ 
are also in the $\Phi^{(n-1)}$ background. 
If we assume that this iterative procedure converges, in the $n\to \infty$
limit we can replace
\be
\Phi^{(n-1)} \to \Phi^{(n)}, \ \ 
\langle 0| \phi^2 |0\rangle^{(n-1)} \to \langle 0| \phi^2 |0\rangle^{(n)} 
\ee
in which case we obtain a closed set of equations
\be
\square \Phi^{(\infty)} + V(\Phi^{(\infty)} ) 
+ \lambda \langle 0| \phi^2 |0\rangle^{(\infty)} \Phi^{(\infty)} = 0
\ee
and the Heisenberg equations for $\phi^{(\infty)}$ are also in the background
of $\Phi^{(\infty)}$. These equations are precisely the CQC equations. Hence
the CQC equations are the semiclassical equations in the infinite iteration limit.

In closing we remark that there is a large literature on applications of the
semiclassical approximation (see {\it e.g.}~\cite{HU1977217,Hu:1978zd,PhysRevD.40.456,PhysRevLett.67.2427}) but we are not aware of any application that deals with fully general backgrounds and 
employs more than one iteration. In addition, the solution for $Z$
determines the quantum operator $\phi$ completely and hence all correlations
functions of $\phi$ can be written in terms of $Z$ and ${\dot Z}$.

\bibliographystyle{JHEP.bst}
\bibliography{coupledCQC}

\end{document}